\documentclass[acmsmall]{acmart}
\AtBeginDocument{%
  }

\setcopyright{acmlicensed}
\copyrightyear{2025}
\acmYear{2025}
\acmDOI{XXXXXXX.XXXXXXX}

\begin{document}

\title{Impact of AI on Software Engineering Jobs}
\title{Will AI Replace Software Engineers? Do not hold your breath}

\author{Abhik Roychoudhury}
\authornote{Abhik Roychoudhury is also a Senior Advisor at SonarSource.}
\email{abhik@comp.nus.edu.sg}
\affiliation{%
  \institution{National University of Singapore}
  \city{Singapore}
  \country{Singapore}
}

\author{Andreas Zeller}
\email{zeller@cispa.de}
\affiliation{%
  \institution{CISPA Helmholtz Center for Information Security}
  \city{Saarbrücken}
  \country{Germany}
}

\renewcommand{\shortauthors}{Roychoudhury and Zeller}

\begin{abstract}

Artificial Intelligence (AI) technology such as Large Language Models (LLMs) have become extremely popular in creating code. This has led to the conjecture that future software jobs will be exclusively conducted by LLMs, and the software industry will cease to exist. But software engineering is much more than producing code---notably, \emph{maintaining} large software and keeping it reliable is a major part of software engineering, which LLMs are not yet capable of.

\end{abstract}



\maketitle

\section{LLMs in Coding}

Large Language Models (LLMs) have shown surprising capability in automatic programming~\cite{ai25}. This includes automated code generation from natural language descriptions and automated remediation of code issues~\cite{acr24,tosem-article}.
It is conjectured that LLMs will take the role of software professionals in the future. But will they? There are significant discussions on this front~\cite{kangshaw24}.

In this paper, let us clarify two common misunderstandings.
First, software engineering is much more than producing program code. At least half of the effort in software engineering is about \emph{maintaining} and \emph{understanding} existing software.
Second, code is more than just text---it has a formal syntax, it has semantics, and it can be executed, tested, and analyzed. This is where the effort goes: understanding what a piece of code does can require lots of domain- and program-specific knowledge; changing it without potentially violating some implicit assumptions can be a daunting task, compromising functionality, security, or safety properties. Learning these assumptions cannot be acquired uniquely from syntactically processing the code, as LLMs do during training; you need to understand the code's execution semantics. And when you track down a bug, you will have to observe code executions and experiment with it. In all these tasks, LLMs can offer some help, but in no way replace human software engineers.


While LLMs are suitable for pattern-based software engineering tasks, the capability of LLMs to capture and reason about \emph{program semantics} remains unknown.  This leads to a  viewpoint that most noncritical software engineering will be automated in the future, with critical systems still involving developers to help reason about code.
The opposite viewpoint could be to prevent LLMs from being used, at least for tasks involving program reasoning.
Both of these viewpoints are extreme. The truth is somewhere in between---and could hold the key to future software job descriptions. This can allow us to understand the evolution of the software professional as we know them today.

\section{"Understanding" Program Semantics}

LLMs excel at syntactic reasoning in a limited context; and, of course, at replicating and adapting solutions for tasks that have been solved before. They are very useful for processing natural language reports. They can produce a first approximation of what the human software engineer would have done, and can reduce the burden. Consider the task of automated program repair, which is supposed to simulate a human engineer, and then provide fix suggestions.  Indeed, LLM agents have been suggested for conducting bug fixing and feature addition, e.g.~\cite{acr24}.

However, as it comes to reasoning about dynamic program execution semantics, LLMs remain thoroughly limited \cite{xia25}. At first glance, it may appear that the recent \emph{reasoning models} such as OpenAI's \texttt{o1} \cite{o1} or \texttt{o3-mini}, 
could be used to ``reason'' about code and thereby decipher program semantics.
These reasoning models employ a long internal chain of thought before answering questions, thereby allowing them to plan the answer with impressive explicit reasoning steps. But this does not indicate their capability to reason about the \emph{dynamic execution semantics} of the code. The fundamental challenge is that with each new line of code, the possibilities in execution multiply, and it takes humans (and automated tools) lots of effort to abstract over all these possibilities. Potential causes and effects in execution are spread all across the code \cite{zeller}, possibly over millions of lines of code. It is unclear how LLMs could capture, explore, and combine all these semantic relations, let alone reason about them without additional help. Many of these problems have exponential complexity or are even undecidable, regardless of the machinery one throws at them.

One alternative could be to train machine learning models not (only) from code sequences but from code \emph{executions.} 
In general, it is useful to extract properties from (a collection of) program executions. These properties can capture control and data flows, as well as more customized input-output relational specifications.
For instance, during code executions, we could collect execution features such as ``Branch \texttt{X} was (not) taken'', ``Variable \texttt{Y} had a value of \texttt{val}'', ``Function \texttt{f()} was (not) reached'', and so on, as well as features of input and output. If we can now generate a multitude of diverse executions (say, via an input generator), training a model on these features would allow it to discover and leverage correlations between them.
Such \textit{program-specific models} could then answer maintenance-related questions like 
\begin{itemize}
    \item ``Where can that value have come from?'',
    \item ``What input do I need to reach this function?''
    \item ``What change to the configuration (input) is needed to make this (output) button appear green?''
\end{itemize}
and more. Training such models might be expensive, but (in contrast to LLMs on code) each of their suggestions can be validated and refined by executing the code. Recent work has shown that accurate models relating input to output features, making predictions in both ways, can actually be learned for small programs~\cite{mammadov2024learningprogrambehavioralmodels}; how far this scales to real-world software and complex executions (and at which cost) remains to be seen.


So while machine learning on \emph{tokens} may not directly reason about dynamic execution semantics, we may be able to use machine learning to observe and summarize patterns from \emph{program executions}, thus supporting understanding, maintenance, and debugging. Program-specific models may then support LLMs in inferring developer intent and thus produce code and repairs that maintain these implicit assumptions as well as explicit requirements~\cite{req24},
both of which are crucial for engineering reliable software.

\section{Perspectives}

LLMs by themselves do not have the capability to take over the software industry in the future. This is because of fundamental limitations in the way they work---limitations that are crucial for safe and secure software development.
Most importantly, directly applying LLMs to any or all software processes, such as those involving reasoning about  dynamic program semantics, should be avoided. Such tasks will require assistance from analysis tools (typically in the form of agents~\cite{aise25}),  findings from dynamic executions, as well as formal/informal input from humans. 


Finally, even when LLMs are used wisely for software tasks, a software development team's cooperative intelligence and diversity of thought will be hard to replicate. We expect LLMs to work as aids or even team members in human software development teams in the future, and they will be welcome additions to the team. Yet, the future software industry will not be a farm of LLMs---not until machines have found how to understand all the intricacies of program code and how to reason about it. 
Given that code already today is among the most complex artifacts ever created by humans, machine-generated code will be no simpler.
And then, if something catastrophic happens with a code base thus produced, who will be there to understand or to fix it?




\begin{acks}
The authors acknowledge discussions with attendees at the IFIP Working Group 2.4 meeting held February 24-27, 2025. This work is partially supported by a Singapore Ministry of Education (MoE) Tier 3 grant MOE-MOET32021-0001. David Lo and Rohan Padhye provided useful pointers.
\end{acks}

\bibliographystyle{ACM-Reference-Format}
\bibliography{main}

\end{document}